\documentclass[sigconf]{acmart}
\usepackage[font=small,labelfont=bf,skip=0pt]{caption}
\usepackage{enumerate}
\usepackage{graphicx}
\usepackage{lipsum}
\usepackage{subfig}
\usepackage{multirow}
\usepackage{tabularx}
\usepackage{mathtools}
\usepackage{balance}
\usepackage{bbm}
\usepackage{color}
\usepackage{float}
\usepackage{listings}
\usepackage{color}
\usepackage{booktabs}
\usepackage{tablefootnote}
\usepackage{url}
\usepackage{threeparttable}
\usepackage{pifont}
\usepackage{tcolorbox}

\definecolor{ao(english)}{rgb}{0.0, 0.5, 0.0}

\newcommand\possiblebreak{\ifhmode\unskip\space\hfil\penalty0\hfilneg\fi}

\usepackage{adjustbox}
\renewcommand{\paragraph}[1]{\textbf{#1} }

\usepackage{algorithmic}
\usepackage[ruled,vlined]{algorithm2e}
\SetAlFnt{\small}

\setcopyright{none}
\renewcommand\footnotetextcopyrightpermission[1]{} 
\begin{document}
\sloppy
\title{A Comprehensive Review and Systematic Analysis of Artificial Intelligence Regulation Policies}
\author{Weiyue~Wu, Shaoshan~Liu}
\maketitle

\section{Abstract}
\label{sec:abs}
Due to the cultural and governance differences of countries around the world, there currently exists a wide spectrum of AI regulation policy proposals that have created a chaos in the global AI regulatory space. Properly regulating AI technologies is extremely challenging, as it requires a delicate balance between legal restrictions and technological developments. In this article, we first present a comprehensive review of AI regulation proposals from different geographical locations and cultural backgrounds. Then, drawing from historical lessons, we develop a framework to facilitate a thorough analysis of AI regulation proposals. Finally, we perform a systematic analysis of these AI regulation proposals to understand how each proposal may fail.  This study, containing historical lessons and analysis methods, aims to help governing bodies untangling the AI regulatory chaos through a divide-and-conquer manner.

\section{Introduction}
\label{sec:intro}

The rapid development of AI technologies in the past decade has spurred the development of many exciting AI services and products. However, due to the cultural and governance differences of countries around the world, there currently exists a wide spectrum of AI regulation policy proposals that have created a chaos in the global AI regulatory space ~\cite{wu2021dilemma}. 

Properly regulating AI technologies is extremely challenging, as it requires a delicate balance between legal restrictions and technological developments. When legal restrictions get ahead of technological developments, the progress of society may be constrained.  Conversely, when technological developments run ahead of legal restrictions, unregulated new technologies may harm human society, defying technological development's fundamental purpose. 

We believe a crucial step of addressing today's AI regulatory chaos is to first understand different AI regulation proposals, and then to systematically analyze each proposal's pros and cons.  This approach provides the necessary apparatus for different governing bodies to gradually evolve their AI regulation policies to maximize the benefits and to minimize the potential hazards of AI technologies. Through the AI regulation evolution process, we hope that different governing bodies can eventually reach consensus towards a global standardized AI regulation framework, which will greatly accelerate the integration of AI technologies into human society. 

Hence, a comprehensive review and systematic analysis of AI regulation proposals is imperative to bring people to a common ground of understanding different AI regulation approaches, and this is exactly what this article aims to achieve. Particularly, we make the following contributions in this article. 

\begin{itemize}
    \item \textbf{A Comprehensive Review of AI Regulation Proposals:} in section \ref{sec:review}, we start by summarizing the AI regulation proposals from different geological are as and cultural backgrounds, including the United States (U.S.), the European Union (EU), the United Kingdom (UK), China, as well as international regulation proposals.  Through this review, we clarify the rationale and approach each country utilizes for AI regulation.
    \item \textbf{A Common Framework for Understanding AI Regulation Proposals:} without a well-constructed evaluation and categorization framework, analyzing AI regulation proposals leads no where. In section \ref{sec:analysis}, we develop a framework to facilitate a thorough analysis of the existing AI regulation proposals, enabling us to evaluate and categorize each proposal in a systematic and objective manner.
    \item \textbf{A Systematic Analysis of AI regulation Proposals:} in section \ref{sec:map}, we combine the learning from section \ref{sec:review} and section \ref{sec:analysis} to analyze and compare each AI regulation proposal, particularly to understand how each existing AI regulation proposal may fail. 
\end{itemize}

\section{Review of AI Regulation Proposals}
\label{sec:review}

The recent rapid development of AI technologies, as exemplified by ChatGPT, has drawn the general public's attention to the capabilities of AI products, pushing AI practitioners to keep pace with, and establish regulations for, these AI products. With the development of AI also come along diverse AI regulation proposals. In this section we compile and compare the AI regulation proposals from different geological and cultural backgrounds aiming for different regulatory rationales.  

In the examination of various AI regulatory models, we introduce two key terms: \textit{Rationale} and \textit{Approach}. Rationale refers to the guiding principle that dictates the focus of regulation. It sets the boundaries of regulatory intervention by identifying which aspects of AI technology are critical for scrutiny and control. On the other hand, approach addresses the execution methods of these regulations. It looks at the structural design of the regulatory system, the assignment of roles among regulatory bodies, and the mechanisms through which rules are enforced. These two terms serve as a structured lens through which different AI regulatory frameworks can be assessed and compared.

\subsection{U.S.’s Industry-Specific Strategy}

While the U.S. has not yet formalized a specific regulatory strategy for AI \cite{horizontallessons}, an examination of historical precedents might offer some insights into possible approaches. AI, in its most fundamental form, comprises data, algorithms, and users. To comprehend the potential regulation of data within AI, it can be instructive to consider how privacy laws, which concern data protection and confidentiality, have been implemented. For algorithm regulation, the principles and methods employed in cybersecurity provide a valuable point of reference, as algorithms are, at their core, software that needs to be secured. Moreover, in understanding how AI users might be protected, we can look to the principles and practices of consumer protection, considering that AI users are essentially the customers.

In each of these domains — privacy, cybersecurity, and consumer protection \textit{the U.S. has always adopted a industry-specific rationale}, as historically regulatory laws within the U.S have been developed in a bottom-up manner such that different industry lobby groups develop initial versions of the regulatory laws and gradually iterate through the legislative branch. In a way, the legislative system in the U.S. has given the liberty to each industry to propose regulatory laws within the industry. 

\textit{The approach the U.S. has taken to enforce regulation also seem industry-specific}. For instance, cybersecurity is a vital concern across all industries, especially as businesses increasingly operate in digital spaces. Cybersecurity is overseen primarily by the cybersecurity and Infrastructure Security Agency (CISA) within the Department of Homeland Security (DHS). However, other agencies like the Federal Trade Commission (FTC) and the Securities and Exchange Commission (SEC) hold industry-specific responsibilities, such as penalizing companies that fail to protect consumer data in the case of FTC, or regulating cybersecurity disclosures in the financial industry in the case of SEC. Similarly, consumer protection is handled across multiple industries. While the FTC is the main consumer protection agency, other agencies, like the Consumer Financial Protection Bureau (CFPB) and the Food and Drug Administration (FDA), play important roles within specific industries.

Given the historical inclination towards industry-specific regulations across these key areas, it is reasonable to assume that the U.S. might adopt a similar industry-specific strategy when it comes to AI regulation. A notable indication of this strategy is reflected in the White House's request for voluntary commitments from leading AI companies to manage AI risks \cite{wh2023}. This request showcases the government's trust in these companies' ability to effectively govern their AI applications.  The advocates for an industry-specific approach argue that professionals who are deeply entrenched in their respective industries possess the most comprehensive understanding of these fields \cite{why2023}. By supplementing these expert teams with AI specialists, the argument goes, a robust and nuanced regulatory framework could be constructed for AI applications within diverse industries.

\subsection{EU’s GDPR-like Strategy}

The EU’s AI Act takes the pattern established by its forerunner, the General Data Protection Regulation (GDPR). The Act \cite{act2023} puts forth an inclusive structure for AI regulation, spanning from laying down requirements for high-risk AI systems to setting up a European Artificial Intelligence Board. The Act prioritizes user safety and fundamental rights, ensures transparency of AI systems, and mandates AI providers to follow strict post-market monitoring rules. This regulatory strategy underscores the EU's commitment to fostering a human-centric and ethical AI landscape, as well as ensuring the broader public's trust in AI technology.

\textit{The rationale behind the EU's AI Act is risk-based}, classifying AI products into distinct categories each subject to varying levels of regulatory requirements. This rationale assesses the potential harm an AI product could cause and accordingly stipulates the necessary precautions. On one end of the spectrum, AI systems and products deemed low-risk, such as spam filters or video game algorithms, may face minimal regulatory restrictions, preserving innovation and usability. Conversely, at the high-risk end, AI systems - like biometric identification and critical infrastructure - are subject to extensive obligations, including the establishment of robust risk management systems and strict transparency requirements for users \cite{eu2021regulatory}. The Act aims to protect users and uphold ethical standards, while accommodating the rapid development and diversification of AI technologies.

To realize the risk-based rationale, \textit{the EU's approach is to establish a new centralized regulatory body} , the Committee on Artificial Intelligence\cite{coe2023}. The key task of this body would be to explicate the legal framework for AI, clarifying and enforcing the regulations set out in the AI Act. The Committee would assume the responsibility for interpreting the specifics of the AI Act, ensuring uniform application across the union, and supervising high-risk AI systems. This approach illustrates the EU's commitment to fostering a comprehensive, coherent regulatory environment, capable of addressing the complexities and potential risks posed by AI technology.

The implementation of EU’s AI Act could face challenges similar to those encountered by its predecessor, GDPR. In a one-year review report, GDPR was criticized for generating unintended consequences rather than effectively protecting privacy \cite{chivot2019evidence}. 

The risk-based rationale might be too simplistic when applied to the intricate realities of AI products. It glosses over the inherent ambiguity and the wide spectrum of risk scenarios associated with AI systems, which do not neatly fall into distinct risk categories. This oversight is highlighted in a recent study \cite{risk2023}, which determined that 18\% of AI systems unambiguously belong to the "high-risk" category, with an additional 40\% potentially qualifying for the same classification. This considerably surpasses the EU's earlier estimates of 5-15\% \cite{impact2023}, underscoring the potential shortcomings of a risk-based rationale.

When it comes to the centralized Committee approach, while it might seem effective on the surface, it could face challenges regarding efficiency and scalability. Given the rapid pace of AI development and the global scope of its deployment, a single, centralized regulatory body may struggle to stay abreast of and effectively address the diverse and quickly-evolving issues associated with AI. Furthermore, potential bottlenecks in decision-making and bureaucratic inertia might impede the swift responses often required in the fast-paced world of AI. Ultimately, while the intent behind the establishment of a centralized Committee on Artificial Intelligence is commendable, its effectiveness in the face of real-world complexities remains to be seen.

\subsection{UK’s Pro-Innovation Strategy}

The UK’s recent policy paper, highlighting a "pro-innovation" strategy to AI regulation, underlines its drive to promote and adapt to the rapid advancement of AI technology \cite{uk2023}. Rather than explicitly labeling it as a "wait-and-see" approach, it's perhaps the accurate term to describe it as an adaptive strategy. At the current stage, the strategy offers non-legally binding guidance, assigning regulatory responsibilities to existing entities, such as the Competition and Markets Authority. It acts as a mechanism for collecting feedback and insights from various stakeholders. This strategy, as per the government's stated goal, aims to enable informed, future decision-making while fostering an environment conducive to innovation and growth in the AI sector.

The UK government introduces the novel concept of a \textit{context-specific rationale}: while U.S. and EU proposals focus on AI products, such as autonomous driving vehicles or facial recognition cameras, the UK's context-specific rationale adds another dimension and evaluates the context in which an AI product is used. For instance, an autonomous vehicle operating in a remote area, like the Scottish Highlands, could potentially face fewer regulations. However, if the same vehicle operates in downtown London, it would be subject to more stringent regulations. This context-specific rationale is believed to foster a nuanced understanding of the deployment and potential risks associated with an AI application, thereby providing room for greater innovation and flexibility in diverse settings.

The regulatory approach in the UK, which we can term the \textit{"contingency" approach}, reflects its distinct handling of the dynamic field of AI governance. The UK has initially dispensed with the idea of establishing a new, AI-focused regulatory body. Yet it acknowledges the necessity for central support mechanisms to manage risk and encourage innovation. This approach is "contingent" because the UK is seeking to balance the need for regulation with the rapidly advancing field of AI. While it has laid out a broad vision, the specifics of execution remain vague, suggesting an ongoing assessment of the situation before making any concrete determinations.

While the UK labels its AI regulatory strategy as "pro-innovation", the reality might become the opposite. An pro-innovation environment necessitates predictable, standardized, and transparent regulations \cite{wu2021dilemma}, enabling entrepreneurs, venture capitalists, and academics to foresee regulatory risks, thereby facilitating rapid assessment and mitigation, and minimizing regulatory uncertainties and emergencies. 

The current UK's AI regulatory proposal lacks these elements, both in rationale and approach. The context-specific rationale, by focusing on myriad potential scenarios of AI applications, may deviate towards a case-by-case analysis, leading to increased discretion and unpredictability. Given the expansive nature of AI, encompassing countless scenarios and products, it's nearly impossible to catalogue all possibilities. This ambiguity makes it difficult to establish predictable, standardized, and transparent regulations and may impede the creation of a truly pro-innovation environment. 

Furthermore, the contingency approach, being largely reactive and subject to change, doesn't foster a pro-innovation environment either. It lacks firm guidelines and concrete decision-making, which might stifle innovation by introducing regulatory uncertainties. The potential unpredictability introduced by this approach might inadvertently deter innovation, contrary to the UK's stated intent.

\subsection{China’s State-Controlled AI Development}

China is positioning AI as a fundamental infrastructure of the Chinese society, seeking to exert state control in the interest of safety and efficiency ~\cite{liu2023change}. This approach mirrors China's historical practice of limiting competition in critical infrastructure sectors, such as energy and electricity \cite{liu2023soe}. State control over AI ensures that private sectors cannot monopolize vital resources, safeguarding national security. Moreover, it enables the concentration of monetary and human capital on the most profitable areas, thereby driving performance. This commitment to regulating AI underlines China's intent to manage its transformation into an essential, ubiquitous element of society.

The recent regulations on generative AI technology demonstrates this intent. The newly implemented framework aligns closely with the principles outlined in the Cybersecurity Law \cite{carnegieendowment2023}. Much like the guardian roles that the Cybersecurity Law assigns to Internet Service Providers (ISPs) and social media platforms, this framework delegates similar oversight responsibilities to generative AI service providers. For example, generative AI service providers are required to terminate services, archive records, and report illegal content to administrative authorities \cite{CAC2023}, a procedure reflective of the one outlined in the cybersecurity law \cite{digichina2017}. Moreover, the speed at which the generative AI regulations \cite{gaichina2023} was rolled out - initiated for consultation in April 2023, only five months after the public release of ChatGPT - reveals China's determined approach to maintaining a tight rein over AI technology.

\textit{China's regulatory rationale and approach towards generative AI are firmly rooted in strict state control}. The authorities' rationale hinges on the idea that tight control can help coordinate the rapid advancement of technology with the need for regulation. This control essentially means that the government's active involvement in AI development and deployment will uphold safety, ensure responsible use, and align the technology's progression with the nation's strategic objectives. As for the regulatory approach, the government's confidence in state control's effectiveness and efficiency is palpable, as evidenced by China's pioneering role in implementing specific regulations for generative AI \cite{cnn2023}.

While the state-controlled approach might appear as a robust strategy, it is important to recognize that the inherent features of AI could make it vulnerable to significant risks. Unlike traditional sectors like road infrastructure and energy, AI is significantly more intertwined with various aspects of society. This integration means that any weakness within the AI infrastructure could potentially escalate into a major crisis at an alarming pace. A recent incident of personal data leakage reinforces this concern: sensitive data from 48.5 million people, collected via a health QR code system used to trace COVID-19 close contacts and suspected cases, was allegedly hacked \cite{healthcodeleak}. Thus, while centralized control of AI promises many benefits, it is also potentially filled with substantial risks that must be vigilantly managed.

\subsection{Establishment of an International Agency for AI Regulation}

In the era of globalization, technological boundaries are clearly fading, it would be greatly beneficial to have a unified, international approach to AI regulation, potentially bridging the cultural and policy differences that may exist between individual countries. Recently, many have raised the concept of establishing an international AI Agency to coordinate global AI regulation \cite{axiossam2023, unsg2023, econ2023}.

The rationale of the proposed International AI Agency is driven by the concept of acting as an \textit{intelligence hub}, a centralized source for shared knowledge and information in the realm of global AI regulation. The approach is tailored towards offering \textit{global standardization}. By harmonizing regulatory standards worldwide, such an international agency aims to provide guidance for nations refining their AI strategies, bridge policy gaps, and foster a cohesive path forward amidst the dynamic progress of AI technology.

One historical example of such an international agency is the International Atomic Energy Agency (IAEA). However, it's crucial to maintain a clear understanding of the distinct nature and challenges presented by AI regulation compared to nuclear regulation. The establishment of IAEA to regulate an industry relies on several pivotal factors. The IAEA's capability to manage nuclear technologies is largely attributed to the controllable number of participants involved. With approximately 440 nuclear power reactors globally ~\cite{reactor2023}, and nuclear weapons confined to only nine countries ~\cite{status2023}, monitoring these entities becomes more manageable. 

Furthermore, the high stakes associated with nuclear technologies, such as the worldwide catastrophe of a nuclear leakage, drive nations to collectively agree on very strict safety protocols. The nuclear industry also benefits from rigid, unified standards that facilitate clear guidance. These conditions, however, are specific to the nuclear industry and may not be transferable to the AI industry, raising questions about the feasibility of an analogous international regulatory body for AI.

AI differs fundamentally from nuclear technologies in all the aspects discussed above, making the establishment of a unified, international regulatory body significantly more complex. First, AI technologies are omnipresent and pervasive, integrated into a myriad of applications across various industries. The sheer volume and diversity of AI applications, from autonomous driving to personalized advertising, present a colossal challenge for any single regulatory body to oversee effectively. Moreover, the implications of AI misuse are multifaceted, ranging from privacy breaches and data theft to the propagation of misinformation and potential job displacement. These varied risks necessitate a broader scope of regulation, which may be difficult to encapsulate in a singular set of international guidelines. Additionally, AI is a rapidly evolving field where standards need to be as dynamic and adaptable as the technology itself. Such flexibility could complicate the creation of a universally accepted regulatory framework.

\subsection{Summary of AI Regulation Proposals}

As highlighted in Table ~\ref{table:sum}, various regulatory proposals are summarized in place, each adopting unique rationales and approaches to address the wide array of concerns brought about by AI. While governing bodies around the world strive to apply historical lessons in managing new technologies, the distinct complexities inherent in AI pose a significant challenge. Considering the rapid evolution and multi-dimensional nature of AI technologies, a more systematic regulatory framework for analyzing AI regulation proposals is imperative.

\begin{table*}[]
\centering
\begin{tabular}{|l|l|l|>{\raggedright\arraybackslash}p{5cm}|>{\raggedright\arraybackslash}p{5cm}|}
\hline
 & \textbf{Rationale} & \textbf{Approach} & \textbf{Benefits} & \textbf{Challenges} \\
\hline
\textbf{U.S.} & Industry-specific & Industry-specific & Preserves existing regulatory frameworks, leverages industry-specific expertise & Risk of inconsistency, overlap and absence \\
\hline
\textbf{EU} & Risk-based & Centralized & Offers a holistic view of AI regulation, sets clear and unified standards & Lack of adaptability and efficiency \\
\hline
\textbf{UK} & Context-specific & Contingent & Recognizes the importance of the application context, allows for flexibility in regulation & Lack of predictability, standardization, and transparency, potential for oversupply of regulatory discretion \\
\hline
\textbf{China} & State-controlled & State-controlled & Coordinate the rapid advancement of technology & Vulnerable to significant risks \\
\hline
\textbf{International} & Intelligence hub & Global standardization & Provides broad applicability & Can be slow to adapt to rapid advancements \\
\hline
\end{tabular}
\vspace{5mm}
\caption{Summary of AI Regulation Proposals}
\label{table:sum}
\end{table*}

\section{A Framework for Analyzing AI Regulation Proposals}
\label{sec:analysis}

Reflecting on the complexity and multi-dimensional challenges posed by AI regulation, it becomes apparent that a systematic methodology is required for dissecting and analyzing the diverse regulatory strategies. In this section, drawing on historical lessons, we develop a framework for a comprehensive analysis of the existing AI regulation proposals.

\subsection{Technological Uncertainty and Regulatory Methods}

Although the emergence of AI represents a unique challenge of technology regulation, it is not the first time human society faces uncertainty in overseeing new technologies. Drawing upon historical lessons can provide valuable insights and possibly pave the way forward.

\begin{figure}
    \centering
    \includegraphics[width=\linewidth]{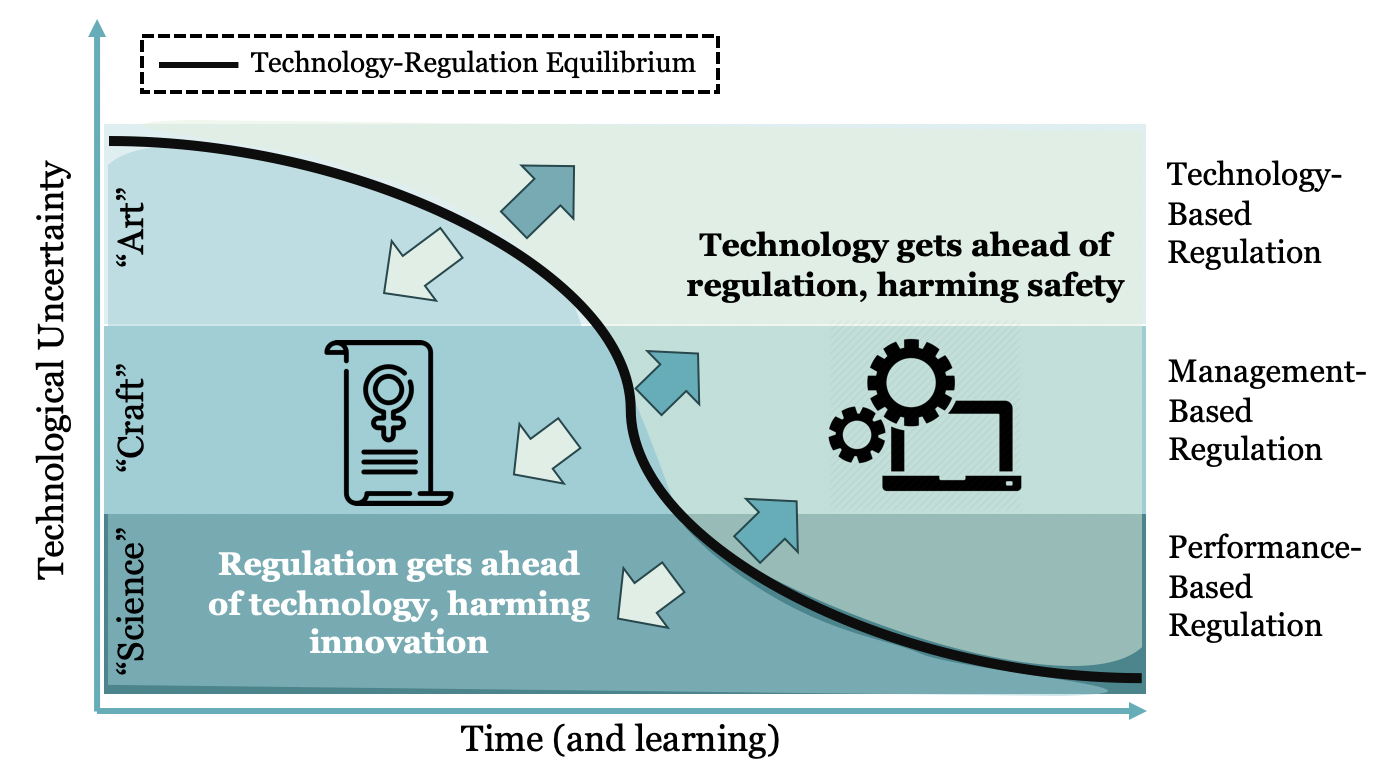}
    \caption{Technological Uncertainty and Regulatory Methods}
    \vspace{5pt}
    \label{fig:frame}
\end{figure}

As shown in Fig. \ref{fig:frame}, a general technology regulation framework provides a useful starting point to understand the interaction between technology uncertainty and the adaptability of regulation ~\cite{roca2017risks}. This framework delineates how regulatory methods can be applied to technologies at different developmental stages. 

This framework categorizes emerging technologies by their respective uncertainties as "art", "craft" and "science". Higher technological uncertainties might bring great benefits as well potential hazards.  In this framework, "art" signifies the highest level of uncertainty, "science" the lowest, whereas "craft" represents an intermediate stage ~\cite{roca2017risks, bohn2005art}. Each level of technological uncertainty demands a suitable type of regulatory method to shepherd the development and to contain the potential hazard of the respective technology. 

To give a few concrete examples, early aircraft designs can be regarded as a form of "art", without comprehensive aerodynamics theories, engineers have to solely rely on trial-and-error methods such that sometimes a particular design would work, and at other times a plane may crash. Internet Search Engine Optimization can be regarded as a form of "craft", as engineers have mastered certain strategies to improve the search engine ranking of a web page, without knowing the algorithmic details of how the search engine works. Semiconductor manufacturing can be regarded as a form of "science", as engineers have deep understanding of of the underlying theories and principles and have full control of the whole process.

For each category of technological uncertainty, different regulatory measures can be applied:

\begin{itemize}
    \item Technology-based Regulation, aligning with the "art" phase, involves imposing rigid technological standards to fulfill regulatory objectives. For instance, achieving a 90\% reduction in automobile tailpipe emissions seemed impossible and a form of "art" in the early 1970s \cite{cleanairact2022}, due to the lack of theory and technologies to control chemical emission at that time. The Environmental Protection Agency (EPA) took a technology-based regulation by making it mandatory for all vehicles to install catalytic converters \cite{epa2023}. This regulation policy quickly enabled the industry to meet the ambitious emissions targets \cite{techforcing}.

    \item Management-based Regulation, harmonizing with the "craft" phase, embodies a regulatory strategy that transfers the decision-making process to those possessing the most relevant information, \textit{e.g.} leading companies in a emerging technology industry. The California Department of Motor Vehicles (DMV) provides an insightful example of the management-based approach in its handling of testing and deployment procedures for autonomous vehicles in 2016. Autonomous vehicles, at that time, could be classified as a form of "craft" in the sense that though a significant amount of operational data and understanding had been accumulated, autonomous vehicles remained largely unpredictable. Rather than imposing rigid technological standards, the California DMV chose a management-based approach centering around the establishment of certain operational and safety conditions based on the suggestions of autonomous driving companies \cite{ dmv2016}. Autonomous driving companies are responsible for reporting any accidents or unplanned disengagements during testing and for resolving these issues prior to deployment. 

    \item Performance-based Regulation, relating to the "science" phase, guides the results while providing manufacturers the leeway to select their preferred solutions. This model fosters innovation by not prescribing the means to achieve the regulatory objective, but only the desired end-state. A case in point is the mandate for Electronic Stability Control (ESC) systems by the National Highway Traffic Safety Administration (NHTSA). Starting in 2012, all passenger vehicles sold in the United States were required to have ESC systems. These systems were expected to meet certain performance criteria under a variety of driving conditions\cite{dot2006}. Importantly, the regulation did not dictate the exact technology or methodology to be used, leaving manufacturers the freedom to choose their own  solutions. 
\end{itemize}

Building upon the discussion above, a technology regulation policy should aim for reaching the equilibrium between technological uncertainty and regulatory restrictions (illustrated by the black line in Fig. ~\ref{fig:frame}). Two scenarios can lead to a deviation from this equilibrium. In the first scenario, represented by a movement towards the top-right direction, indicating that technological uncertainty has surpassed regulatory restrictions, potentially leading to unbridled technological growth without necessary constraints. The state of radio frequency technology before 1912 illustrates the potentially chaotic outcomes of unregulated technological uncertainty. The absence of a formalized regulatory framework for allocating radio frequencies resulted in rampant interference among amateur, maritime, and transoceanic communications. This disarray of communications, largely attributable to the lack of regulation, significantly contributed to the tragic sinking of the Titanic \cite{hoolihan2016radio}. 

Conversely, a shift towards the bottom-left direction signifies that technological developments are overly restricted, inhibiting the healthy development of technologies. One example has been reported by an AI startup company, indicating that due to overly strict regulatory requirements, they had to spend over 40\% of their budget on regulatory compliance, compared to 15\% of the software industry norm ~\cite{wu2023compliance}.

\subsection{Analysis Framework}
 
\begin{figure}
    \centering
    \includegraphics[width=\linewidth]{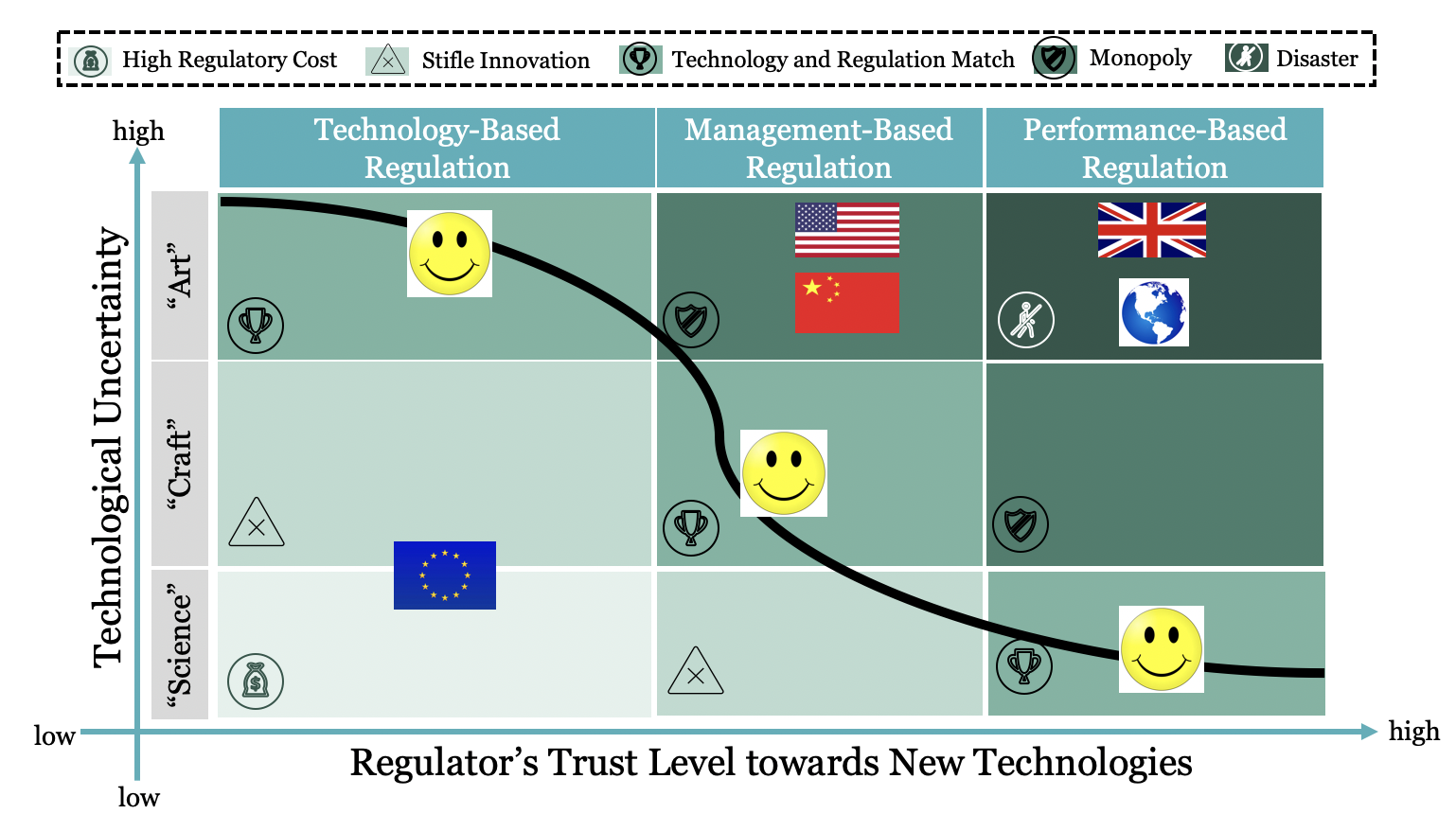}
    \caption{Matrix of Technology Regulation}
    \vspace{5pt}
    \label{fig:matrix}
\end{figure}

In this subsection, we develop a framework for evaluating various AI regulation proposals (Fig. ~\ref{fig:matrix}). The proposed framework couples the three levels of technological uncertainty with three corresponding regulatory methods, resulting in nine distinct boxes.  

We categorize these boxes into five distinct situations to explain the consequences of a match or mismatch of technological uncertainty and regulatory methods (annotated in Fig. ~\ref{fig:matrix}). 

\paragraph{Technology and Regulation Match:} The optimal scenario where regulatory methods aligns with technological uncertainty. This balance fosters an environment conducive to innovation and safety. This is illustrated by the black line in Fig. ~\ref{fig:matrix}.

\paragraph{Monopoly:} When applying management-based regulation to technologies still in the "art" phase, \textit{i.e.}, relying on major companies to regulate new technologies that are under development by them, this would allow companies to monopolize the market due to their regulatory and lobbying influence, and developing rules in favor of the companies' interests. 

\paragraph{Disaster:} The extreme upper right box represents a scenario where technology drastically outstrips regulation, or applying performance-based regulation on technologies still in the "art" phase. The consequences of this situation could be catastrophic. Performance-based regulation depends heavily on universal technological standards. When technological uncertainty is high and such universal standards are immature, performance-based regulation may essentially be tantamount to an absence of regulation.

\paragraph{Stifle Innovation:} Shifting towards the lower left, the box represents a scenario where regulation significantly outpaces technology development. This can occur when technology-based regulation is applied to "craft" level technologies, or when management-based regulation is applied to "science" level technologies.  For instance, currently, the autonomous driving technology sits at the "craft" phase where there is considerable experimentation and varying approaches to achieving full autonomy. If regulators were to dictate specific sensor configurations or AI algorithms, it could stifle the innovative solutions being pursued by different companies and constrain the potential advancements in this field.

\paragraph{High Regulatory Cost:} Technology-based approach becomes increasingly counter-productive when applied to "science" phase technologies, primarily due to the technology's maturity and broad consensus around operational standards. Adhering to specific technological requirements can be incredibly time-consuming and costly, as it fails to leverage the standardization and flexibility offered by performance-based regulation.

\section{How Existing AI Regulation Proposals May Fail}
\label{sec:map}

In this section, we utilize the framework developed in section \ref{sec:analysis} to evaluate the AI regulation proposals summarized in section \ref{sec:review} and explore how each proposal may fail. The categorization of each AI regulation proposal is illustrated in Fig. \ref{fig:matrix}.

\subsection{U.S.: Management-Based Regulation May Lead to Monopoly or Oligopoly}

The U.S. is showing tendencies towards adopting a management-based approach to regulating AI. The White House has conducted meetings with CEOs from tech giants including Google, Microsoft, OpenAI, and Anthropic to stress the responsibility of these companies in the development of safe AI technologies ~\cite{house2023readout}. This signifies a strategy where the regulators are relying on these tech companies to manage and uphold standards within their own organizations, which aligns with the tradition of technology regulation within the U.S. A key drawback of the management-based regulation approach is the leading to monopoly or oligopoly, this has already happened in most U.S industries where only a few leading companies dominate the whole market ~\cite{foge2019american}. 

AI is universal and monopolistic in nature, none of the technologies that came before AI had such a widespread impact on human society, and only the best AI engine will dominate the whole market.  Consequently, AI is not going to impact one or two industries, but likely to dominate across all industries. Aware of the widespread and profound impact of AI and potential unreadiness of the U.S. AI regulation policy, some scholars have established an AI Incident Database, inviting the general public to report suspicious AI incidents \cite{incidentdatabase2023}. This grassroots effort aims to increase awareness of the potential risks associated with AI across all industries.

We can project that if management-based regulation leads to monopoly or oligopoly in AI technologies, then very soon this may result in a few giant tech firms monopolizing the whole U.S. economy. Indeed, the big tech domination of the U.S economy is already happening, as indicated by U.S. Senator Marco Rubio in a recent speech \cite{rubio2023}.

\subsection{China: State Monopoly of AI}

Unlike in the U.S. where management-based regulation may lead to a few tech giants dominating the whole U.S economy, the Chinese state may purposely aim to monopolize AI technologies through its state-owned enterprises (SOEs) \cite{liu2023state, liu2023soe}. In this setup, the Chinese state act as both the regulator and the supplier of AI technologies. 

Until a few decades ago, China had always been an agricultural nation, such that for thousands of years the core production resource was land. There existed a clear social contract between the state and the people, where the state kept the people under control through land allocation, and in return the people cultivated the lands and gave back a fraction of their production in the form of taxes to sustain the state’s operation. 

In the past few decades, as China transitioned from an agrarian society to an industrial powerhouse and then to a service society, the state also imposed tight controls on production resources such as energy, telecommunications, and transportation through state-owned enterprises (SOEs). In this setup, the state ensured its control over the provisioning and controlling of critical production resources, as well as generating profits indirectly through the SOEs. But the social contract between the state and the people remains the same: The state provides production resources; the people cultivate these resources to make a living and pay taxes to the state. 

Hence, if the same social contract persists, instead of having the people cultivating land provided by the state and paying taxes to the state, the new generation of Chinese people will cultivate AI models provided by the state and pay taxes to the state in return. From this perspective, China’s government will definitely monopolize AI, potentially through a new SOE.

However, this approach also presents significant challenges and raises a number of important questions. For instance, how can regulators ensure that these tech companies are meeting their obligations if both of them are owned by the state? How transparent will these companies be about their internal AI management systems? What mechanisms will be in place to handle non-compliance? Without the clear separation of regulation and technology development, these questions can not be answered.

\subsection{EU: Stifle Innovation and High Cost}

With its risk-based rationale, EU is applying technology-based management to regulate AI, which may stifle innovation and impose high compliance costs for companies developing AI technologies, restricting the healthy development of AI and the integration of AI technologies into society.   

EU's risk-based regulatory strategy for AI comes with its own unique challenges, chief among which is the delineation of boundaries between each risk category. The EU's approach is premised on categorizing AI technologies based on the level of risk they present, from low to high. However, the complexity and dynamism inherent in AI technologies can often blur these boundaries, making the categorization process fraught with ambiguity and uncertainty.

Firstly, AI systems are not static. They often learn and evolve over time, such that the level of risk an AI system presents can change throughout its lifecycle. This fluidity can complicate efforts to categorize a given AI product into a static risk category, thus posing challenges to consistent regulatory enforcement.

Secondly, different AI applications can have vastly different impacts depending on their use context. For example, facial recognition technology could be low risk when used in a personal smartphone, but high risk when used for mass surveillance. Thus, the task of categorizing AI systems by risk level involves not only understanding the technology itself, but also the varied contexts in which it may be deployed. This adds another layer of complexity and potential ambiguity to the process.

In such a volatile landscape, where the level of risk associated with an AI system can shift frequently and unexpectedly, stakeholders may struggle to adapt. Companies may find it difficult to predict the regulatory implications of their AI products, leading to uncertainty in product development and investment decisions. Similarly, regulators may struggle to keep up with the rapid pace of AI development and its shifting risk profiles, which could impede effective oversight.

Thus, while the EU's risk-based approach attempts to be flexible and adaptive to the diverse nature of AI technologies, it requires clear, comprehensive guidelines for risk categorization and a robust mechanism for continuous monitoring and reassessment. Without these, the ambiguity in AI product categorization could become a significant stumbling block in its effective implementation.

\subsection{Performance-Based Management May Lead to Catastrophic Results}

Both the UK and the international AI agency proposals fall into the category of applying performance-based management for AI regulation. However, AI technologies are still in infancy with unpredictable development paths and outcomes. Having lenient regulatory restrictions such as performance-based management may lead to catastrophic results. 

The UK takes a pro-innovation approach towards emerging AI technologies and apply a performance-based regulation approach. Here, rather than stipulating how things must be done (technology-based) or who does them (management-based), the focus is on the outcome. In such a situation, companies are typically given the maximum flexibility to choose their preferred methods to achieve the desired performance outcome. This can lead to a host of unintended and potentially catastrophic consequences such as violation of privacy, ethical breaches, unfair decision-making, and even systematic failures if these AI technologies are widely adopted. The Facebook-Cambridge Analytica data scandal of 2018 \cite{guardian2019}, which significantly interfered with the UK Brexit referendum, stands as a vivid illustration of the potential unintended consequences when companies are given extensive flexibility in achieving their desired outcomes. 

Similar to the UK case, the proposal for an international AI Agency can be regarded as a form of performance-based regulation as it aims to moderate the outcome and allows maximum flexibility for the developers in determining their methods of achieving the desired results. The issue stems from AI's high degree of uncertainty and diversity, and the lack of universally accepted standards or benchmarks to measure AI's performance. This lack of clarity and standardization makes it challenging for a single regulatory body to effectively oversee and moderate the outcomes of a wide spectrum of AI applications.

\section{Summary}
\label{sec:sum}


Although the human society has developed very sophisticated systems for technology regulation since the Industrial Revolution, exemplified by internet laws, none of the technologies that came before AI had such a widespread impact on human society in such an early stage of its development. Because AI's impact is so widespread, having universal strict regulations may impede the development of many sectors. Because AI is still in its infancy, letting it grow freely may lead to disasters. Hence, regulating AI requires a delicate balance between legal restrictions and technological developments.

Moving forward, AI is too complex to apply one rule on, as AI is not one technology, but a broad technical term that contains many facades. From a foundation model perspective, AI can be regarded as a form of “art”, similar to early aircraft design, in a way that the engineers solely rely on trial-and-error. From an application development perspective, AI can be regarded as a form of “craft”, similar to Internet Search Engine Optimization, in a way that the engineers have mastered certain strategies to improve a page's ranking on search engines but do not fully understand the details of algorithms powering these search engines. For certain AI application usage scenarios, such as the usage of a Face ID, AI may even be regarded as a form of "science", similar to semiconductor manufacturing, in a way that the engineers have deep understanding and fine control over every details of the process.

The framework developed in this article aims to help governing bodies untangling the AI regulatory chaos through a divide-and-conquer approach, first properly categorizing a particular AI regulation case, then applying the appropriate regulatory measures potentially drawing lessons from historical examples, and finally continuously evaluating and refining the regulatory policies.

\bibliographystyle{ieeetr}
\footnotesize{%
\bibliography{ref}

\begin{thebibliography}{10}

\bibitem{wu2021dilemma}
W.~Wu and S.~Liu, ``Dilemma of the artificial intelligence regulatory
  landscape,'' {\em Communications of the ACM}.

\bibitem{horizontallessons}
{MATT O'SHAUGHNESSY,  MATT SHEEHAN}, ``Lessons from world's two experiments
  in ai governance,'' February 2023.

\bibitem{wh2023}
{The White House}, ``Fact sheet: Biden-harris administration secures voluntary
  commitments from leading artificial intelligence companies to manage the
  risks posed by ai.''
  https://www.whitehouse.gov/briefing-room/statements-releases/2023/07/21/fact-sheet-biden-harris-administration-secures-voluntary-commitments-from-leading-artificial-intelligence-companies-to-manage-the-risks-posed-by-ai/.
\newblock Accessed: 2023-07-20.

\bibitem{why2023}
``Why a single ai regulator just won’t work.''
  \url{https://thehill.com/opinion/congress-blog/4020159-why-a-single-ai-regulator-just-wont-work/}.
\newblock Accessed: 2023-07-05.

\bibitem{act2023}
``Ai act: a step closer to the first rules on artificial intelligence.''
  \url{https://www.europarl.europa.eu/pdfs/news/expert/2023/5/press_release/20230505IPR84904/20230505IPR84904_en.pdf}.
\newblock Accessed: 2023-07-05.

\bibitem{eu2021regulatory}
``Eu commission: Regulatory framework proposal on artificial intelligence.''
  \url{https://digital-strategy.ec.europa.eu/en/policies/regulatory-framework-ai}.
\newblock Accessed: 2023-07-05.

\bibitem{coe2023}
``The council of europe \& artificial intelligence.''
  \url{https://rm.coe.int/brochure-artificial-intelligence-en-march-2023-print/1680aab8e6}.
\newblock Accessed: 2023-07-05.

\bibitem{chivot2019evidence}
E.~Chivot, D.~Castro, E.~Chivot, and D.~Castro, ``What the evidence shows about
  the impact of the gdpr after one year,'' {\em Center for Data Innovation},
  2019.

\bibitem{risk2023}
``Ai act: Risk classification of ai systems from a practical perspective.''
  \url{https://aai.frb.io/assets/files/AI-Act-Risk-Classification-Study-appliedAI-March-2023.pdf}.
\newblock Accessed: 2023-07-05.

\bibitem{impact2023}
``Impact assessment of the regulation on artificial intelligence.''
  \url{https://digital-strategy.ec.europa.eu/en/library/impact-assessment-regulation-artificial-intelligence}.
\newblock Accessed: 2023-07-05.

\bibitem{uk2023}
``Ai regulation: a pro-innovation approach.''
  \url{https://www.gov.uk/government/publications/ai-regulation-a-pro-innovation-approach}.
\newblock Accessed: 2023-07-05.

\bibitem{liu2023change}
S.~Liu, ``Artificial intelligence will bring social changes in china,'' {\em
  the Diplomat}, 2023.

\bibitem{liu2023soe}
S.~Liu, ``Will china create a new state-owned enterprise to monopolize
  artificial intelligence?,'' {\em the Diplomat}, 2023.

\bibitem{carnegieendowment2023}
``China's ai regulations and how they get made.''
  \url{https://carnegieendowment.org/2023/07/10/china-s-ai-regulations-and-how-they-get-made-pub-90117}.
\newblock Accessed: 2023-07-05.

\bibitem{CAC2023}
``Measures for the management of generative artificial intelligence services.''
  \url{http://www.cac.gov.cn/2023-07/13/c_1690898327029107.htm}.
\newblock Accessed: 2023-07-05.

\bibitem{digichina2017}
``Translation: Cybersecurity law of the people's republic of china (effective
  june 1, 2017).''
  \url{https://digichina.stanford.edu/work/translation-cybersecurity-law-of-the-peoples-republic-of-china-effective-june-1-2017/}.
\newblock Accessed: 2023-07-05.

\bibitem{gaichina2023}
``Translation: Measures for the management of generative artificial
  intelligence services (draft for comment) – april 2023.''
  \url{https://digichina.stanford.edu/work/translation-measures-for-the-management-of-generative-artificial-intelligence-services-draft-for-comment-april-2023/}.
\newblock Accessed: 2023-07-05.

\bibitem{cnn2023}
``China unveils new ai regulations to address ethical concerns.''
  https://edition.cnn.com/2023/07/14/tech/china-ai-regulation-intl-hnk/index.html.
\newblock Accessed: 2023-07-05.

\bibitem{healthcodeleak}
``Shanghai health code system containing personal data of 485 million.''
  https://www.scmp.com/tech/big-tech/article/3188746/shanghai-health-code-system-containing-personal-data-485-million.
\newblock Accessed: 2023-07-05.

\bibitem{axiossam2023}
``Ai leaders: Sam altman and others call for regulation in senate testimony.''
  \url{https://www.axios.com/2023/05/17/ai-leaders-sam-altman-regulate-senate}.
\newblock Accessed: 2023-07-05.

\bibitem{unsg2023}
``Secretary-general urges broad engagement from all stakeholders towards united
  nations code of conduct for information integrity on digital platforms.''
  https://press.un.org/en/2023/sgsm21832.doc.htm.
\newblock Accessed: 2023-07-05.

\bibitem{econ2023}
``The world needs an international agency for artificial intelligence, say two
  ai experts.''
  \url{https://www.economist.com/by-invitation/2023/04/18/the-world-needs-an-international-agency-for-artificial-intelligence-say-two-ai-experts}.
\newblock Accessed: 2023-07-05.

\bibitem{reactor2023}
``Plans for new reactors worldwide.''
  \url{https://world-nuclear.org/information-library/current-and-future-generation/plans-for-new-reactors-worldwide.aspx}.
\newblock Accessed: 2023-07-05.

\bibitem{status2023}
``Status of world nuclear forces.''
  \url{https://fas.org/initiative/status-world-nuclear-forces/}.
\newblock Accessed: 2023-07-05.

\bibitem{roca2017risks}
J.~B. Roca, P.~Vaishnav, M.~G. Morgan, J.~Mendon{\c{c}}a, and E.~Fuchs, ``When
  risks cannot be seen: Regulating uncertainty in emerging technologies,'' {\em
  Research Policy}, vol.~46, no.~7, pp.~1215--1233, 2017.

\bibitem{bohn2005art}
R.~E. Bohn {\em et~al.}, ``From art to science in manufacturing: The evolution
  of technological knowledge,'' {\em Foundations and Trends{\textregistered} in
  Technology, Information and Operations Management}, vol.~1, no.~2, pp.~1--82,
  2005.

\bibitem{cleanairact2022}
``Clean air act: A summary of the act and its major requirements.''
  https://crsreports.congress.gov/product/pdf/RL/RL30853.
\newblock Accessed: 2023-07-05.

\bibitem{epa2023}
``Timeline of major accomplishments in transportation air pollution and climate
  change.''
  https://www.epa.gov/transportation-air-pollution-and-climate-change/timeline-major-accomplishments-transportation-air.
\newblock Accessed: 2023-07-05.

\bibitem{techforcing}
D.~Gerard and L.~B. Lave, ``Implementing technology-forcing policies: The 1970
  clean air act amendments and the introduction of advanced automotive
  emissions controls in the united states,'' {\em Technological Forecasting and
  Social Change}, vol.~72, no.~7, pp.~761--778, 2005.

\bibitem{dmv2016}
``Department of motor vehicles: Invitation to pre-notice public discussions on
  proposed regulations autonomous vehicles.''
  https://www.dmv.ca.gov/portal/file/autonomous-vehicles-workshop-notice-september-2016-pdf/.
\newblock Accessed: 2023-07-05.

\bibitem{dot2006}
``{DEPARTMENT OF TRANSPORTATION National Highway Traffic Safety Administration
  49 CFR Parts 571 and 585}.'' Regulatory Document, 2006.
\newblock [Docket No. NHTSA–2006-25801] [RIN: 2127-AJ77].

\bibitem{hoolihan2016radio}
D.~D. Hoolihan, ``The radio act of 1912,'' {\em IEEE Electromagnetic
  Compatibility Magazine}, vol.~5, no.~2, pp.~32--34, 2016.

\bibitem{wu2023compliance}
W.~Wu and S.~Liu, ``Why compliance costs of ai commercialization maybe holding
  start-ups back,'' {\em Harvard Kennedy School Review}.

\bibitem{house2023readout}
W.~House, ``Readout of white house meeting with ceos on advancing responsible
  artificial intelligence innovation,'' {\em The White House}, 2023.

\bibitem{foge2019american}
Z.~Foge, ``American oligarchy: How the enfeebling of antitrust law corrodes the
  republic,'' {\em J. Bus. Entrepreneurship \& L.}, vol.~12, p.~119, 2019.

\bibitem{incidentdatabase2023}
``Ai incident database.'' https://incidentdatabase.ai/about/.
\newblock Accessed: 2023-07-20.

\bibitem{rubio2023}
``Rubio: A change in direction will define our success or failure in the 21st
  century.'' \url{https://www.youtube.com/watch?v=anmxlydLDuc&t=13s}.
\newblock Accessed: 2023-07-05.

\bibitem{liu2023state}
S.~Liu, ``The future of state-sponsored ai research in china,'' {\em the
  Diplomat}, 2023.

\bibitem{guardian2019}
``The cambridge analytica scandal changed the world, but it didn't change
  facebook.''
  https://www.theguardian.com/technology/2019/mar/17/the-cambridge-analytica-scandal-changed-the-world-but-it-didnt-change-facebook.
\newblock Accessed: 2023-07-05.

\end{thebibliography}
}
\end{document}